**Letter**

# In vivo localization of Fas-associated death domain protein in the nucleus and cytoplasm of normal thyroid and liver cells

**Tourneur, Léa[1,2], Alain Schmitt[1,2], and Gilles Chiocchia[1, 2, 3]**

From Inserm, U567, Institut Cochin, Paris, France[1].

Université Paris Descartes, CNRS (UMR 8104), Paris, France[2.]

Service de rhumatologie, Hôpital Ambroise Paré, Boulogne, France[3].

Address all correspondence and requests for reprints to: Dr. Gilles Chiocchia, Institut Cochin, Département d'Immunologie, INSERM U 567, 27 rue du Faubourg Saint-Jacques, 75674 Paris Cedex 14, France. Phone: (33) 1 40 51 66 15. Fax: (33) 1 40 51 66 41. E-mail: gilles.chiocchia@inserm.fr.

**Abstract**

FADD (Fas-associated death domain) is the main death receptor adaptor molecule that transmits apoptotic signal. Recently, FADD protein was shown to be expressed both in the cytoplasm and nucleus of in vitro cell lines. In contrast to the cytoplasmic FADD, the nuclear FADD was shown to protect cells from apoptosis. However, in vivo subcellular localization of FADD was still unknown. Here, we demonstrated that FADD protein was expressed in both cytoplasmic and nuclear compartment in ex vivo thyroid cells demonstrating that nuclear sublocalization of FADD protein was a relevant phenomenon occurring in vivo. Moreover, we showed that in the nucleus of untransformed thyroid cells FADD localized mainly on euchromatin. We confirmed the nuclear localization of FADD in ex vivo liver and showed that in this organ FADD and MBD4 interact together. These results demonstrate that FADD is physiologically expressed in the nucleus of cells in at least two mouse organs. This particular localization opens new possible role of FADD in vivo either as an inhibitor of cell death, or as a transcription factor, or as a molecular link between apoptosis and genome surveillance.

**Introduction**

Normal thyrocytes constitutively express the Fas death receptor but not the Fas ligand (FasL) [1-3]. However, FasL expression in thyroid follicular cells (TFC) can occur under numerous thyroid pathological conditions, including autoimmune diseases and cancer [4-6]. For instance, FasL expression has been reported in TFC from patients suffering of Graves' disease [7] and Hashimoto's thyroiditis [8], the two main thyroid autoimmune disorders in humans, and thyroid carcinomas [6]. In these pathologies, thyrocytes appear to be relatively insensitive to Fas-mediated cell death although expressed FasL is functional [9-12]. Reasons for such phenomenon are not clearly established, and probably involve several mechanisms such as expression of FLIP (FLICE-inhibitory protein) and Bcl-xL anti-apoptotic molecules [11], or loss of FADD (Fas-associated death domain) pro-apoptotic protein [13]. Interestingly, normal TFC are naturally resistant to Fas-induced apoptosis suggesting that the observed high insensitivity of pathological thyrocytes could result from an intrinsic property of normal thyroid cells. Since FADD physically interacts with death receptors located at the cell membrane, FADD protein was thought to be mainly cytoplasmic. Recently, we reported by mean of an in vitro organ culture the existence of a new regulatory mechanism of FADD protein expression following adenosine receptor signaling [14].

Furthermore, the human FADD protein contains both a nuclear export and a nuclear localization sequence which account for FADD localization both in the cytoplasm and the nucleus, respectively [15]. Whereas cytoplasmic FADD possesses pro-apoptotic functions, it was recently reported that FADD expression in the nucleus protects cells from apoptosis [15]. The mechanism implicated in this survival function has not been investigated.

As dogmas can be challenged, FADD nuclear sublocalization has been debated. It was found that FADD primarily resided in the nucleus of cells and thereafter shuttled from the nucleus to the cytoplasm [16]. In contrast, other report showed that FADD was exclusively located in the cytoplasm of most adherent and non adherent cell lines [17]. It now appears that the discrepancies between these two studies likely resulted from different technical approaches [18, 19].

Although it is now demonstrated that FADD is expressed in the nucleus of both human and mouse cell lines [20], and in the nucleus of human lung adenocarcinoma cells [21], no data are available concerning the FADD subcellular localization in non transformed cell in vivo. Thus, one important issue that needed to be clarified was the in vivo FADD subcellular localization. Regarding the innate resistance of TFC to apoptosis, and the important role of the Fas pathway in thyroid autoimmune and tumoral diseases, we addressed this issue in ex vivo mouse thyroid sections. We demonstrated for the first time that nuclear localization of FADD protein is not a cell line dependent event. We confirmed this result in mouse liver and showed that FADD is linked to MBD4 in vivo.

**Materials and methods**

Mice

Different strains of mice (CBA/J, DBA/1, DBA/2, C57BL/6 or BALB/c mice) (Iffa Credo, L'Arbresle, France, and Harlan Olac, Bicester, GB) were used at 7-15 weeks of age. All mice were maintained in standard environmental conditions, and allowed to adapt to their environment at least for one week before the experiments. The studies were approved by the Cochin institute committee on animal care. Agreement number to perform experiments on living animals: n° 75-777, and animal facility agreement number n° 3991.

*Immunofluorescence*

Animals were sacrificed and the two thyroid lobes removed. Collected lobes were immediately covered in optimal temperature medium (Tissue-Tek; Bayer, Elkhart, IN), slowly frozen by floating in isopentane on liquid nitrogen, and stored at -80°C until use. Sections of 5–6 $\mu$m were cut on a cryostat at -18°C and collected onto SuperFrostplus slides (Roth Sochiel, Lauterbourg, France). Sections were dried overnight and stored at -80°C until use. Before staining, sections were fixed for 15 min in PBS with 2% paraformaldehyde (PFA) at 4°C, and incubated for 30 min in PBS with 2% bovine serum albumin. Then, sections were stained (60 min) with primary goat polyclonal IgG anti-mouse FADD antibody (10 $\mu$g/ml, clone M19, Santa Cruz Biotechnology, TebuBio, Le Perray en Yvelines, France) or with isotype-matched control antibody at the same concentration (Vector Laboratories, AbCys, Paris, France). The secondary biotin-conjugated anti-goat IgG antibody (Vector Laboratories) was used at 1 $\mu$g/ml (30 min). Alexa Fluor® 488 conjugate streptavidin (10 $\mu$g/ml, Molecular Probes Inc., Interchim, Montluçon, France) was used to visualize specific staining (30 min incubation at room temperature and protected from light). After washings in PBS, sections were mounted in VECTASHIELD® Mounting Medium with DAPI (4',6-diamidino-2-phenylindole) (Vector Laboratories) to counter-stain DNA. Cells were analyzed by confocal fluorescence microscopy (Bio-Rad MRC1024, Bio-Rad Laboratories) equipped with a digital Diaphot 300 system. Digital pictures were analyzed using LaserSharp software and processed using Adobe Photoshop®.

*Immunogold electron microscopy*

Thyroid lobes were fixed in 1% glutaraldehyde in 0.1 M phosphate buffer pH 7.4, then embedded in sucrose and frozen in liquid nitrogen. Cryosections were made using an ultracryomicrotome (Reichert Ultracut S.) and ultrathin sections mounted on Formvar-coated nickel grids were prepared. Briefly, the sections were incubated for 15 min with PBS 15% glycine, for 5 min with PBS 15% glycine 0.1% BSA and for 20 min with PBS 15% glycine, 0.1% BSA, 10% normal donkey serum followed by 2 h incubation with the goat polyclonal anti-FADD antibody M19. The primary antibodies were diluted to 2 $\mu$g/ml in PBS 15% glycine, 0.1% BSA, 4% normal donkey serum. After extensive rinsing in PBS 15% glycine, 0.1% BSA, sections were incubated for 1 h with gold-labeled secondary rabbit anti-goat antibody with a gold particle size of 10 nm (GAM 10, British Biocell, Cardiff, Wales). Sections were then washed for 30 min with PBS 15% glycine, stained with 2% uranyl acetate for 10 min and air dried. Examination was performed with a Philips CM 10 electron microscope.

*Preparation of protein extracts*

Proteins of ex vivo mouse liver were extracted with the Nuclear Extract Kit (Active Motif, Europe, Rixensarf, Belgium) following manufacturer's instructions. Sample concentration was determined using micro BCA protein assay reagent kit (Pierce, Rockford, IL, USA).

*Immunoprecipitation*

The nuclear extracts were precleared with proteins G sepharose (P-3296, Sigma-Aldrich chimie SARL, Saint Quentin Fallavier, France) on rocking for 1 h at 4°C. Then, 1 $\mu$g of either rabbit polyclonal anti-FADD (AB3102, Chemicon International, Temecula, CA, USA), or rabbit polyclonal anti-MBD4 (ab3756, abcam, Cambridge, United Kingdom), or isotype-matched control antibody (normal rabbit IgG, Santa Cruz Biotechnology) was added at the precleared nuclear extracts, and the mix were incubated for 2 h on ice. Thereafter, 16 $\mu$g of clear proteins G sepharose was added. After an additional 2 h on rotating wheel at 4ºC, precipitates were washed five times with 500 $\mu$l of lysis buffer (10 mM Tris HCl pH 7.8, 150 mM NaCl, 1% NP40, a protease cocktail inhibitor set (complete Mini EDTA-Free, Roche, Indianapolis, USA)). Then, immunoprecipitates were analyzed by FADD specific western blotting.

*Western Blot*

40 µg of cytoplasmic or nuclear proteins, and the immunoprecipitates were diluted in reducing sample buffer, subjected to 15 % SDS-PAGE, transferred to PVDF membrane (NEN Life Sciences, Boston, Massachusetts), and probed with specific primary anti-FADD antibody (M-19, 0.2 µg/ml in TTBS 0.1% containing 5% milk) (Santa Cruz Biotechnology) following by peroxidase-conjugated anti-goat IgG (0.66 µg/ml) secondary antibody (Sigma-Aldrich). Proteins were visualized using the enhanced chemiluminescence technique (Amersham Pharmacia Biotech, Orsay, France). Bands obtained were quantified by densitometry using biocapt® and bio-profil bio1d® software. The same amount of nuclear proteins and cytoplasmic proteins were loaded on gel although the total amount of nuclear protein was five to ten times lesser than the amount of cytoplasmic proteins.

**Results and discussion**

Animals were sacrificed and the thyroid removed. The two thyroid lobes were either immediately covered in optimal temperature embedding medium for immunofluorescence analysis (Fig. 1), or immediately fixed with glutaraldehyde for immunogold electron microscopy (Fig. 2). To avoid the problems associated with permeabilisation reagents such as saponin [18, 19], we performed thyroid cryosections that we immunolabeled with the previously characterized specific anti-mouse FADD antibody [13].

Confirming our previous study [13], confocal immunofluorescence microscopy analysis showed that FADD protein was expressed in the cytoplasm of all TFC (Figure 1 A, E, G), and in all thyroid examined independently of the genetic background of the mice (data not shown). Moreover, we observed that FADD was additionally localized in the nucleus of some, but not all, TFC (Fig. 1 A, B, E-G). In contrast, omission of the primary antibody or staining with an isotype-matched control antibody showed no positive reactivity in the nucleus (Fig. 1 C, D, H-J, and data not shown). To confirm these results, and to localize more precisely the FADD protein in TFC, we used immunogold electron microscopy. This method allowed us to detect FADD protein both in the cytoplasm and in the nucleus of TFC (Fig. 2 A and B), confirming the results obtained by confocal microscopy. Furthermore, the electron microscopy technique showed that FADD was expressed in the nuclear compartment of all the TFC examined (data not shown).

The different cell-types in tissue samples provided useful internal comparators, when assessing FADD immunolabelling. We found that FADD protein was not, or barely detected in the nucleus of endothelial cells from blood vessel which are localized between thyroid follicles (Fig. 2 E and F). Since FADD was present in the nucleus of ex vivo mouse thyroid cells, our data demonstrated that nuclear localization of FADD protein was not an in vitro culture cell line dependent event. However, the TFC restricted pattern of expression of nuclear FADD protein in thyroid gland suggested that nuclear sublocalization of this protein may not be a common feature and may depend on cell type. Finally, we confirmed that FADD could be expressed in the nucleus of cells in vivo using a third approach which is cellular fractionation. Because of the large amount of protein needed for such technic, we used proteins of ex vivo mouse liver instead of thyroid. We chose liver to compare with thyroid because both hepatocytes and thyrocytes are epithelial cells. Whatever the amounts of protein extracts loaded on SDS-PAGE, we detected FADD expression both in the cytoplasmic and nuclear fractions (Fig. 3 A). Quantification of bands obtained by western blot showed that the nuclear FADD represents one third of the total FADD protein, demonstrating that in this organ FADD is mainly cytoplasmic (Fig. 3 B). These results confirmed that FADD is present in the nucleus of two types of mouse epithelial cells in vivo.

We observed that part of FADD was localized in the nucleus. This result differed from several results reporting the nuclear localization of only a minor fraction of FADD. On the other hand it was in line with the results reported in adherent cell lines in which FADD seems to primarily reside in the nucleus, and thereafter to translocate to the cytoplasm [16] a phenomenon which depends on phosphorylation at serine 194. Here in we estimated that in our setting around 20% of the total FADD was localized in the nucleus.

FADD is a complex molecule and some of its assigned properties have not been yet fully demonstrated. For instance, the death effector filaments (DEF) structures are cytoplasmic filament network composed of death effector domain (DED)-containing proteins, including FADD [22], that recruit and activate pro-caspase 8 [23]. However, the existence of DEF has not been established in vivo, and increasing evidences suggest that DEF could be structures resulting only from protein over-expression [15, 16]. Indeed, DEF formation has always been observed in the context of FADD protein over-expression, and DEF have been characterized as detergent-insoluble in contrast to endogenous FADD which is soluble in NP40 or triton. Here, we demonstrated that the ability of FADD to localize in the nuclear compartment is a natural phenomenon since it occurred in primary normal cells ex vivo. Moreover, we showed that FADD protein localized in the nucleus of thyrocytes and liver cells in physiologic conditions as well as in the cytoplasm, demonstrating that in vivo FADD does not have exclusive subcellular localization in these two cell types.

The function of nuclear FADD, in contrast to the well established role of cytoplasmic FADD in cell death [24, 25], is poorly understood. Whereas FADD interaction with death receptors activates caspase activation cascade in the cytoplasm, thus leading to the apoptotic death of the cell, nuclear FADD appears to be implicated in survival mechanisms [15]. It has been suggested that sequestration of FADD in the nucleus could protect cells from apoptosis by hindering FADD-triggered caspase interactions in the cytoplasm [16]. Indeed, it has been demonstrated in cell line that the methyl-CpG binding domain protein 4 (MBD4) could interact with FADD in the nucleus. The relative level of FADD, MBD4, and caspase 8 may determine the apoptotic output of the cells (18). Here in, we tested if FADD could interact with MBD4 in vivo. We performed coimmunoprecipitation experiments using rabbit polyclonal anti-MBD4 or anti-FADD antibodies, and revealed the presence of FADD protein by specific western blotting. We could not perform the converse experiment since the anti-MBD4 did not work well for western blot detection. Immunoprecipitates obtained from nuclear extract of liver cells contained a FADD signal, in contrast to those obtained with isotype-matched control rabbit IgG (Fig. 4). The anti-FADD antibody did not allow immunoprecipitation of all the nuclear FADD since the protein was still detected in the supernatant after immunoprecipitation (Fig. 4 A). By contrast, all the nuclear FADD was immunoprecipitated by the anti-MBD4 antibody demonstrated that all the nuclear FADD was bound to MBD4 (Fig. 4 A).

These results demonstrated that FADD interacts with MBD4 in vivo. This interaction by contributing to FADD sequestration in the nucleus could be involved in protection of cells from death receptor-mediated apoptosis. Alternatively, knowing the key role of MBD4 in GT mismatches repair [26], MBD4-FADD interaction could link genome surveillance/DNA repair and apoptosis in liver cells in vivo.

On the other hand, FADD could be a transcription factor which could induce anti-apoptotic gene expression [15]. Interestingly, using immunogold electron microscopy we observed in the nucleus of TFC that the FADD protein localized almost exclusively on euchromatin, i.e.

noncondensed DNA (Fig. 2 B-D). This particular localization was observed in nucleus of all TFC (data not shown). These results are in favor of FADD playing a role as a regulator of transcription in thyroid cells. Now that we have demonstrated the dual subcellular localization of FADD in TFC, further experiments are needed to determine whether nuclear FADD is a real transcription factor and whether it might regulate expression of anti-apoptotic gene, in particular in pathological TFC. This could contribute to the high resistance of TFC to death receptor-mediated cell death despite the concomitant presence of FADD in the cytoplasm [2, 13]. Characterization of potential FADD-DNA binding sequences or FADD-protein binding partners in nucleus might provide keys for this concept.

**Acknowledgements**

The authors would like to thank S Mistou for his great assistance with cell culture and protein studies and M. Garfa for his excellent technical assistance with confocal fluorescence microscopy. They are also indebted to F. Lager for help with animal care. They are grateful to C. Fournier for critical reading of the manuscript. Léa Tourneur was a recipient of a "Société Française d'Hématologie" (SFH) post-doctoral training fellowship. This work was supported by the Institut National de la Santé Et de la Recherche Médicale (INSERM) and by Ligue Nationale Contre le Cancer-Comité de Paris.

**Figures Legend**

Fig. 1. Analysis of FADD subcellular localization by immunofluorescence confocal microscopy. (A, E) E*x* vivo thyroid sections were stained with anti-FADD antibody (green fluorescence). (C, H) Staining of thyroid sections with Alexa Fluor® 488 anti-goat antibody alone showed no positive reactivity. The same result was obtained using isotype-matched control antibody (data not shown). (B, D, F, I) Nucleus of TFC were counter-stained with DAPI (blue fluorescence). (G, J) FADD immunolabelling and DAPI merge. Co: colloid; Cy: cytoplasm; N: nucleus. Bars indicate scale.

Fig. 2. Analysis of FADD subcellular localization by immunogold electron microscopy. FADD (indicated by arrowheads) is found both in the cytoplasm and the nucleus of TFC (A-D) but not in the nucleus of endothelial cells from blood vessel (E, F). C, D, and F represent higher magnification of B and E, respectively (see framed). Staining with secondary anti-goat antibody alone or isotype-matched control antibody showed no immunogold deposition (data not shown). Co: colloid; Cy: cytoplasm; N: nucleus; RBC: red blood cell. Bars indicate scale.

Fig. 3. Analysis of FADD subcellular localization by western blot after cellular fractionation. (A) Western blot was performed using different amount of protein extracts from ex vivo liver, as noted in the figure. (B) Histogram represents quantification of bands obtained in (A). Results are expressed in arbitrary units. Full and empty histograms represent cytoplasmic and nuclear fractions, respectively.

Fig. 4. FADD interacts with MBD4 in vivo. (A) Western blot analysis of FADD in the cytoplasm (C) and nucleus (N) of ex vivo liver. Nucleus lysates was immunoprecipitated (IP)

with isotype-matched control (ctrl), anti-FADD (FADD), or anti-MBD4 (MBD4) antibodies. Immunoprecipitates were analyzed by FADD specific western blotting.

(B) Histogram represents quantification of bands obtained in (A) after normalization to adjust the amount of proteins loaded to 40 $\mu$g. Results are expressed in arbitrary units.

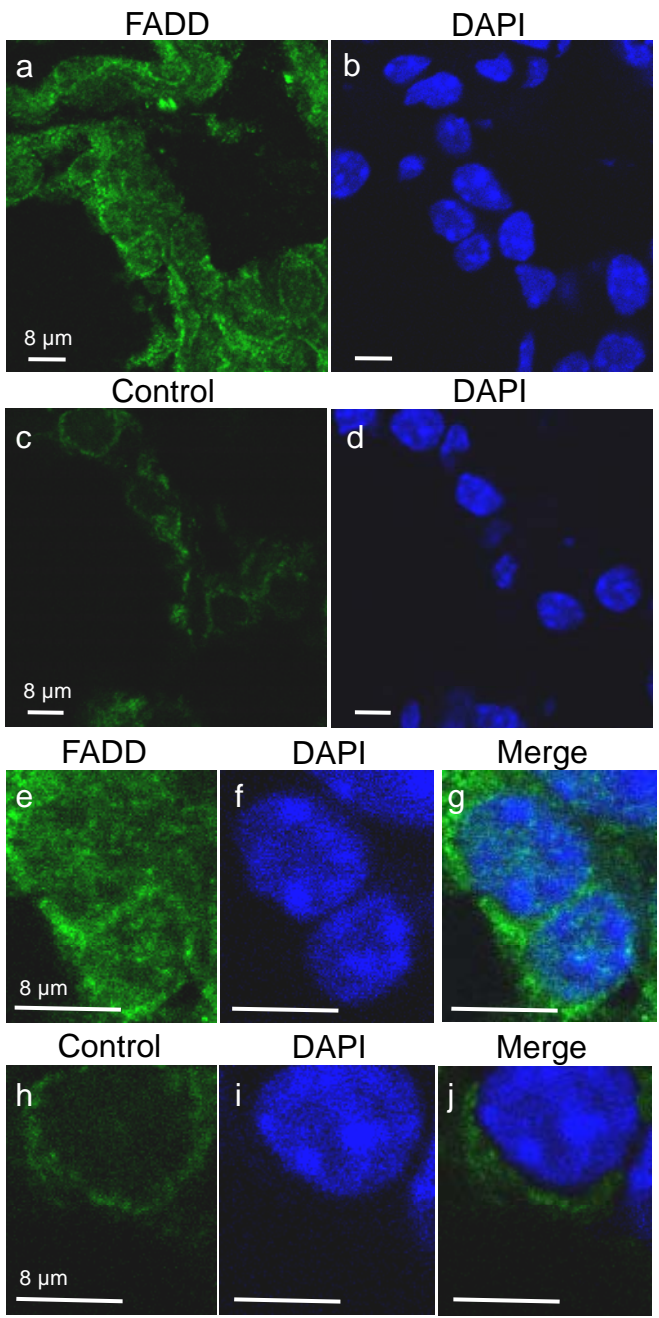

Figure 1

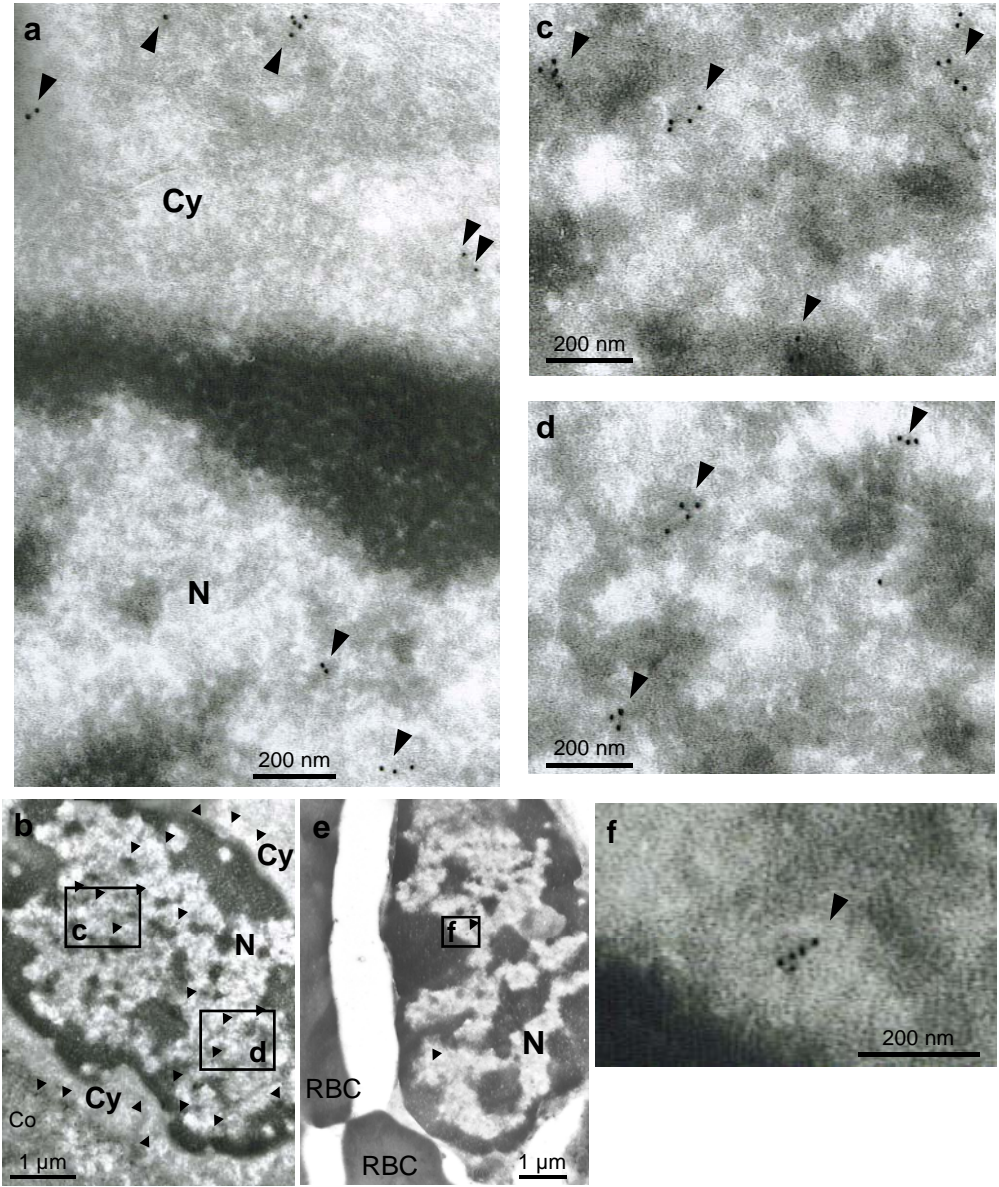

Figure 2

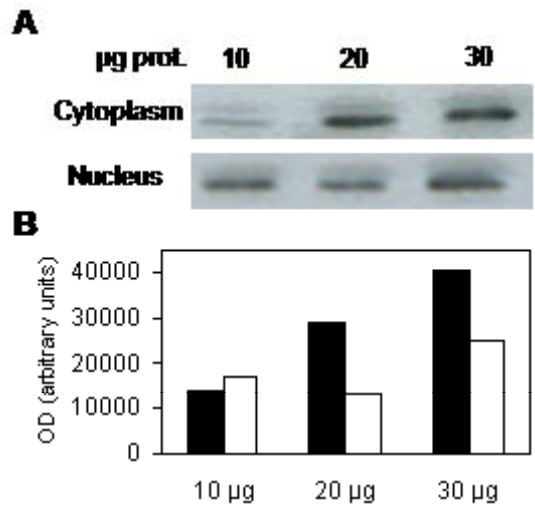

Figure 3

A

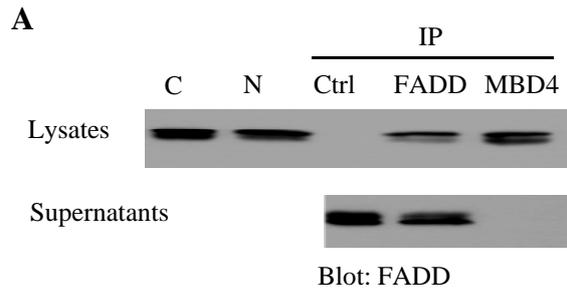

Blot: FADD

B

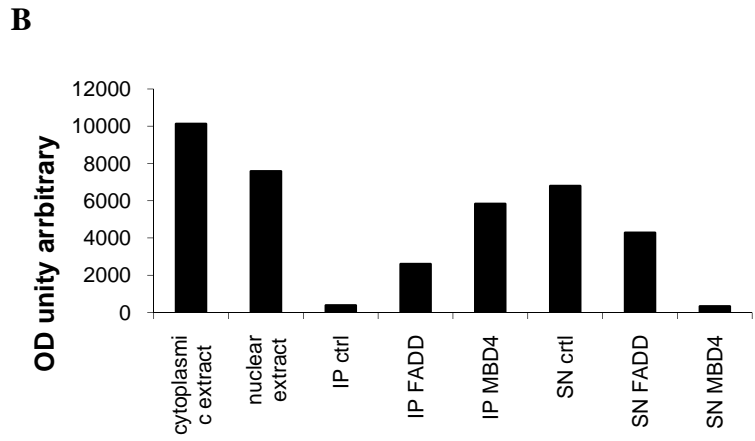